\documentclass[aps,pre,floats,twocolumn,superscriptaddress]{revtex4}

\usepackage{amssymb,amsfonts,amsmath}
\usepackage[pdftex]{graphicx}
\usepackage{multirow}

\begin{document}

\title{Structural efficiency of percolation landscapes in \\flow networks}

\author{M. \'Angeles Serrano}

\author{Paolo De Los Rios}

\affiliation{Institute of Theoretical Physics, LBS, SB, EPFL, 1015
Lausanne, Switzerland}

\begin{abstract}
Complex networks characterized by global transport processes rely on the presence of directed paths from input
to output nodes and edges, which organize in characteristic linked components. The analysis of such network-spanning structures in the framework of percolation theory, and in particular the key role of edge interfaces bridging the communication between core and periphery, allow us to shed light on the structural properties of real and theoretical
flow networks, and to define criteria and quantities to characterize their efficiency at the interplay between structure and functionality. In particular, it is possible to assess that an optimal flow network
should look like a "hairy ball", so to minimize bottleneck effects and the sensitivity to
failures. Moreover, the thorough analysis of two real networks, the Internet customer-provider set of relationships at the autonomous system level and
the nervous system of the worm {\it Caenorhabditis elegans} --that have been shaped by very different dynamics and in
very different time-scales--, reveals that whereas biological evolution has selected
a structure close to the optimal layout, market competition does not necessarily tend toward
the most customer efficient architecture.
\end{abstract}

\maketitle

\section{Introduction}
Despite profound differences, natural and artificial
networked systems share striking similarities. Complex networks
science~\cite{Albert:2002,Dorogovtsev:2003,Newman:2003} has
successfully rationalized several of the most ubiquitous features, such as
the small world property or the presence of strong degree heterogeneity, relating them to the existence of general organizing principles. These self-organization laws also shape the observed large-scale connectivity layout of a special, yet common, class of networks describing transport processes, be it of  matter, energy, or information. These networks are characterized by asymmetric interactions giving rise to local flows that
collectively organize into a large-scale stream dominated by a processing core which transfers input into
output: the universal bow-tie architecture~\cite{Broder:2000} that is intimately related to the functional activity of these systems.

In general terms, most previous research exploring the
relation between form and function in complex networks has been mainly focused on the
analysis of topological features such as modular ordering revealing
functional aspects~\cite{Guimera:2007}, with fewer exceptions treating directly functional aspects such as efficiency~\cite{Latora:2001}. Specifically, transport has been studied as one of the main
functions influenced by topology~\cite{Sreenivasan:2007,Gallos:2007} and functional design principles of global flux distributions have been discussed for biological networks~\cite{Segre:2002,Csete:2004,Fisher:2005}. Despite
these efforts, the ``form follows function'' assertion still remains to be fully understood from a complex
networks science perspective, a major difficulty in the fact that present network patterns are the result of non-stationary and
adaptive evolutionary histories that can greatly vary for different networks. However, general self-organization principles should not only govern structure but also their interplay with functional features.

Our purpose of inferring information about function and evolution from a precise
knowledge of the topological makeup requires the understanding of how flow networks organize to develop functionality.
In this respect, percolation theory on complex networks~\cite{Newman:2001b} provides
a valuable framework to discuss their connectedness and to identify the components that are key to a complete description of their global connectivity layouts conforming the percolation landscapes. This analysis, in turn, allows us to quantify
the degree of efficiency that the network has achieved in relation to its operativeness as a global transport system. In particular,
the major role played by interfaces, bridging the communication between the different percolation components~\cite{Serrano:2007c}, allowed us to define structural efficiency in terms of two complementary aspects: stress or structural load carried by the interfaces --which also informs about robustness--, and closeness or extent of the direct access to the
processing core. We use theoretical arguments to propose the conformation of
maximal structural efficiency and demonstrate by the analysis of
real networks that biological systems exposed to long-term evolutionary
pressure may be much closer to optimality than information technologies systems at an
early stage of development dominated by competitive forces.

\section{The architecture of percolation landscapes}
\begin{figure}[t]
\begin{center}
\includegraphics[width=8.7cm]{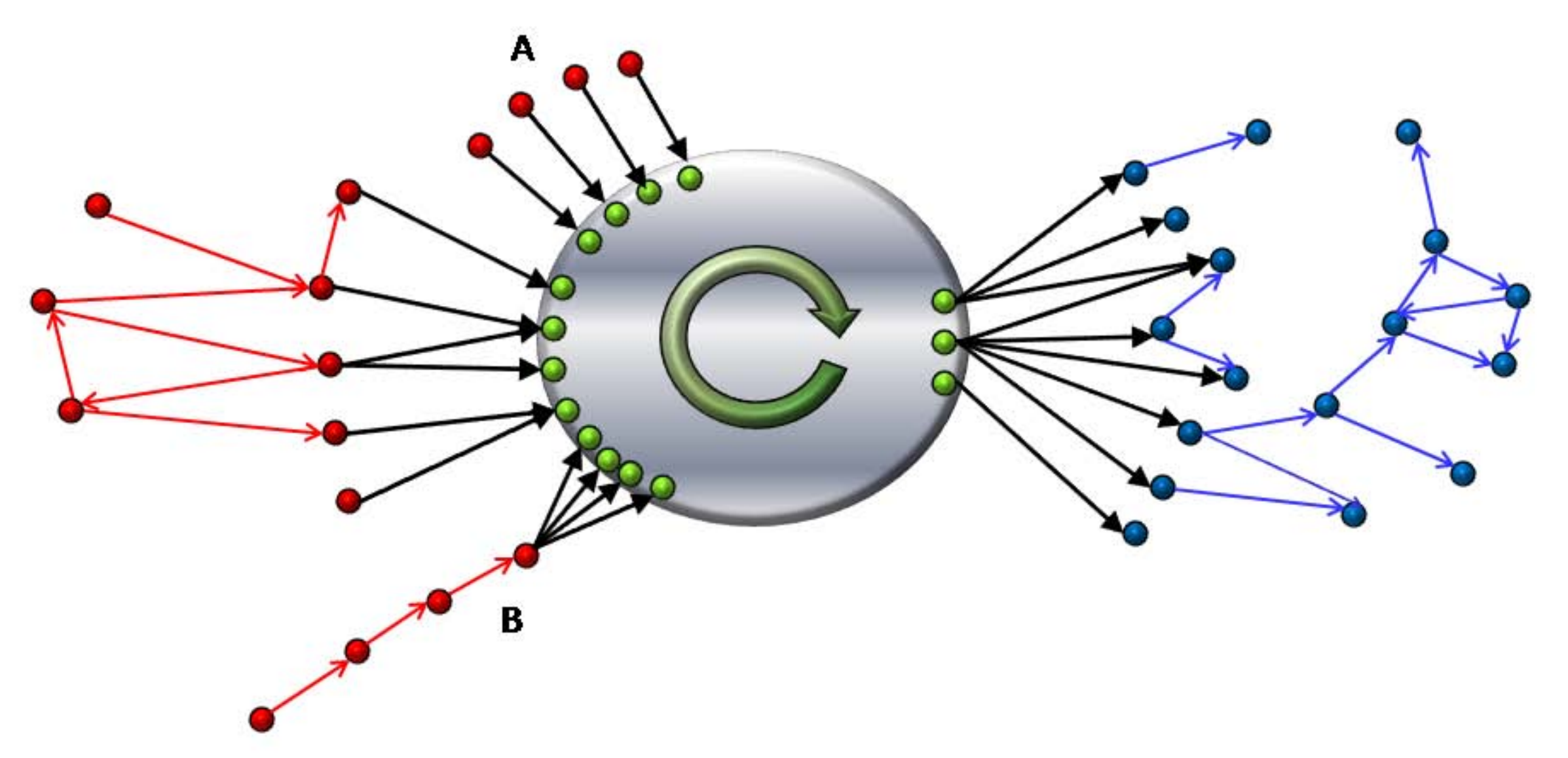}
\vspace{-0.5cm}
\caption{Schematic diagram of the main components in the percolation
landscape of a flow network. The core at the center comprises nodes in the SCC and edges within, forming the SCE. Nodes in red belong to
the IN and the ICE is formed by red links. Nodes in blue belong to
the OUT and blue links form the OCE. Both interfaces, ITF and OTF, appear in
black.} \label{fig:str}
\end{center}
\end{figure}
Global communication is essential to develop efficient collective behavior. In flow networks, represented as directed complex networks, global connectivity is ensured by the presence of architectural elements that allow to traverse the network from the input to the output components. These layouts are best rationalized in the framework of percolation theory, so we call them percolation landscapes. Characteristic
topologies in the percolated phase denote a global flux that organize in distinct linked
components comprising macroscopic portions of the system (Fig.~1 gives a schematic representation). In the percolated phase, the traditional node percolation map~\cite{Newman:2002a,Dorogovtsev:2001,Boguna:2005} recognizes a core structure, the giant strongly connected component (SCC),
which vertices can communicate with each other following directed
paths. In many real systems this core is a processing unit which
transfers input to output, and so it is connected to peripheral
components. The input comes from an afferent component, the giant
in-component (IN), composed by all vertices that can reach the SCC
but cannot be reached from it, and the output goes to an efferent
component, the giant out-component (OUT), made of all vertices that
are reachable from the SCC but cannot reach it. Secondary
structures such as tubes or tendrils could also be present~\cite{Broder:2000}.
Changing
the perspective from nodes to edges, this picture is complemented by
the edge percolation map~\cite{Serrano:2007c}, where the
number of relevant structures increases to five: the edges pure components, ICE, OCE, and SCE, that are formed by edges connecting nodes within the IN, OUT, and SCC respectively; and edges forming the interfaces, ITF and OTF, that bridge the peripheral components (IN and OUT, respectively) to the core.

This pattern is obviously further shaped by system
dependent specificities that are the reflection
of functional demands and evolutionary and/or adaptive
forces. In particular, the specific conformation of the interfaces
determines the structural efficiency and robustness of the
network as a global transport system and the potential risk of bottleneck effects.

\begin{figure}[t]
\begin{center}
\includegraphics[width=8.7cm]{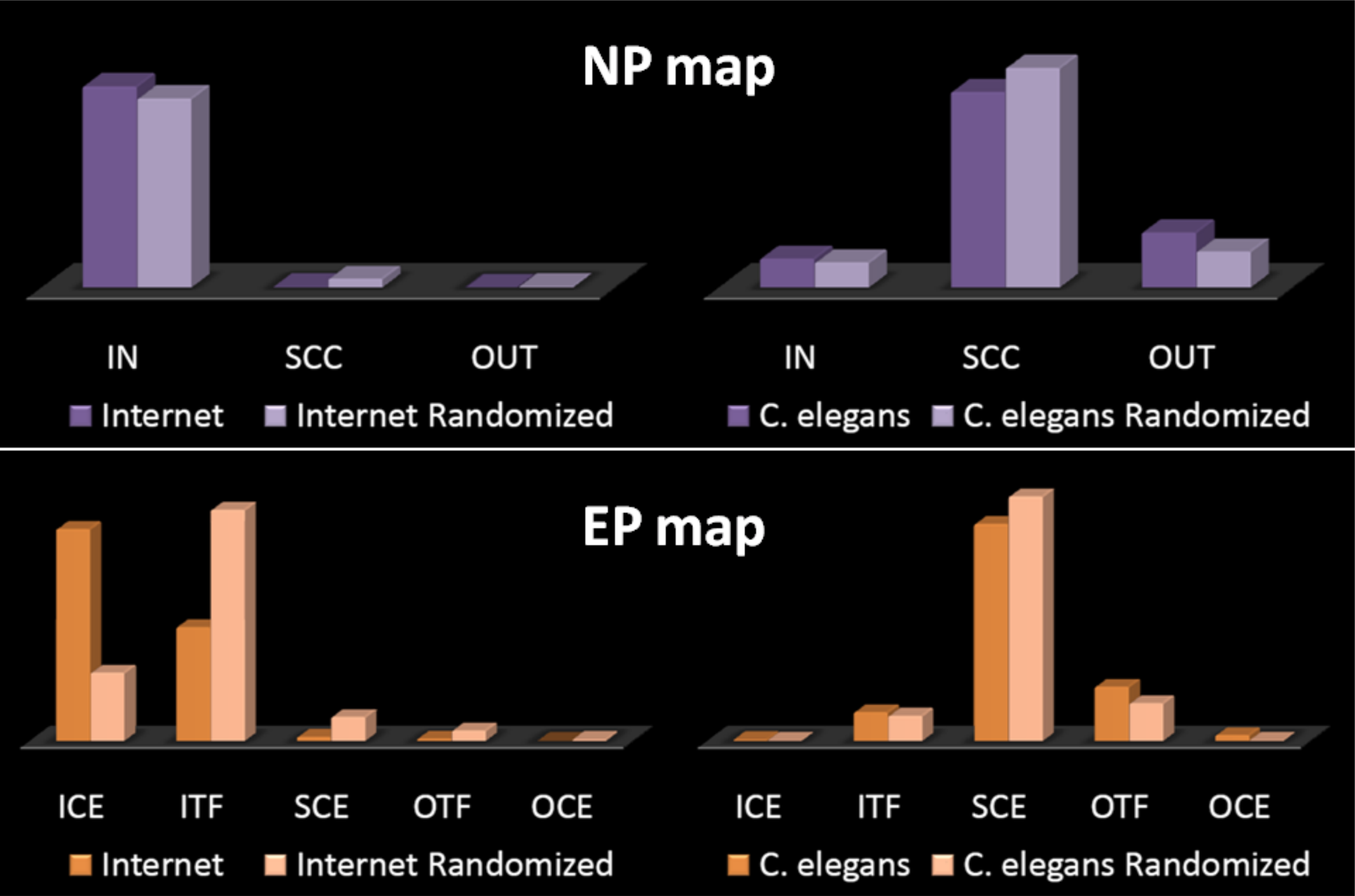}
\vspace{-0.5cm}
\caption{Bar diagrams of the values showed in Table~1 detailing the
percolation landscapes of the ASR and the CEN networks as compared
with the randomized counterparts. Top charts in violet show the node
percolation maps and bottom charts in orange show the edge
percolation maps.} \label{fig:pl}
\end{center}
\end{figure}
\begin{table}[t]
\caption{Statistics for the percolation landscapes of the ASR and
the CEN networks and their randomized versions (subscript $R$, values are average $\pm$ standard deviation rounded off to the first significative figure). The sizes of the main components
are given in absolute number of nodes and links. The average degrees
are $\left<k_i\right>=\left<k_o\right>=1.87$ and
$\left<k_i\right>=\left<k_o\right>=6.82$ respectively. NP stands for
Node Percolation map and EP for Edge Percolation map.}
\begin{center}
\begin{tabular}{llrr@{$\pm$}lrr@{$\pm$}l}
&&ASR&\multicolumn{2}{c}{ASR$_R$} &CEN&\multicolumn{2}{c}{CEN$_R$}\\
\hline \hline
\multirow{3}{*}{NP}&IN&  20060  & 18900&800  & 29    & 25&2 \\
&SCC& 90   & 880&40  &195    &  219&2     \\
&OUT& 17 &  120&30 & 55     &36&2    \\
\cline{2-8}
&Main& 20167   &  19900&800   &  279    & 279&0       \\
&TOTAL& 24545   &  \multicolumn{2}{c}{$\hspace{-0.7cm}$24545}&      279    & \multicolumn{2}{c}{279}   \\
\hline\hline
\multirow{5}{*}{EP}&ICE& 20180   & 6500&400  &9  &3&2  \\
&ITF& 10833  & 22000&2000      &175  &154&8   \\
&SCE& 389   &  2300&200   & 1322     & 1490&20    \\
&OTC&226  & 1000&100     &330  & 230&20     \\
&OCE& 12  & 150&50     &36   & 4&3 \\
\cline{2-8}
&Main& 31640  &32000&2000  &1872      & 1883&5   \\
&TOTAL& 45914   &  \multicolumn{2}{c}{$\hspace{-0.7cm}$45914} & 1903      & \multicolumn{2}{c}{$\hspace{-0.15cm}$1903}   \\
\hline \hline
\end{tabular}
\end{center}
\label{table1}
\end{table}

\subsection{Percolation landscapes of real networks}
We consider here two different information
processing systems characterized by global transport phenomena: one socio-technological, the Internet, in
contraposition with one biological, the nervous system of the
nematode worm {\it Caenorhabditis elegans} (C. elegans).
The node and edge percolation maps of their directed network
reconstructions are detailed in
Fig.~2 and Table~1. 

Their maximally random counterparts are also analyzed as null models. In practice, the randomization is achieved at the
stationary state of a rewiring process that at each time step
randomly selects a couple of links and exchange their ending
points~\cite{Maslov:2002} avoiding the formation of
multiple and self-connections and bidirectional links while preserving the
degree distribution $P(k)$. The randomized version would preserve
as well degree–-degree correlations and higher-order effects which
correspond to structural constraints ensuring the realizability of
the network~\cite{Boguna:2004}. The comparison of real networks with
their randomized counterparts makes therefore possible to determine
to which extent the measured values are due to global organizing
principles and not to random assemblages affected by finite-size
effects. In this work, the randomized counterparts are produced out of $10$ randomized realizations.

\subsubsection{Internet customer-provider AS relationships (ASR)}
The Internet is
one of the paradigmatic information technology and communication
networks~\cite{RomusVespasbook}. From an operative point of
view, it is composed of thousands of Internet Service Providers, usually identified with autonomous systems (ASs), that operate individual parts of the whole infrastructure and engage in contractual relationships to collectively route traffic
through the network. Such business
dependencies~\cite{Dimitropoulos:2007} are
mappable to a directed graph representation of unambiguous customer-provider
relationships among ASs.

The directed graph is reconstructed from the map 2007-04-02 of inferred AS relationships provided by
CAIDA (http://www.caida.org/data/active/as-relationships/).
Relationships among ASs are usually realized in the form of business
agreements, generally simplified to customer-provider, peer-to-peer
and sibling-to-sibling. In a purely directed version of the network, where links represent net flow of payments for services provided,
relations between siblings immediately cancel out since they administratively belong to
the same organization. Peer-to-peer
relations are however not trivial because they just freely exchange traffic between
themselves and their customers but not up in the hierarchy. Anyway, we assume here that the later
are balanced in both directions so as a first approximation we
neglect them as well. On the other hand,
customer-provider relationships are unambiguously represented by
directed edges from customer to provider. We are left with a purely
directed network of $24545$ nodes and $45914$ directed links, after
removing $4312$ ($8.55\%$) peer-to-peer and $236$ ($0.47\%$)
sibling-to-sibling relations). The in-degree distribution is very
broad and well described by a power law with characteristic exponent
$2.1$. The out-degree distribution is strongly bounded and decays extremely
fast with a maximum out degree of $24$.

This network presents an extremely asymmetric
structure at the level of the node percolation map, with
a huge IN component, a restricted SCC, and an even smaller OUT
component (Table~1).
By comparison, the randomized counterpart is characterized by a similar IN component, but by ten-fold larger (albeit still small in absolute terms) SCC and OUT components.
This information about the node partition should be complemented by the analysis of the edge percolation map to provide a first glimpse of the different architectural organization of the real versus the randomized network. Again the size in number of edges of the core and the efferent structures (see left graph in Fig.~3 for details about the efferent components) are qualitatively in accordance with the values for the randomized network, despite being smaller. However, the organization of the afferent components is very different from random. The ICE of the real network contains as many edges as nodes in the IN component. Moreover, the number of ITF edges connecting the IN and SCC components is just half the number of IN nodes: on average, thus, there are two IN nodes for every ITF edge, which further implies that many nodes in the in-component lack direct access to the core. By converse, the randomization predicts an ITF double in number of edges than actually observed with a correspondingly reduced ICE, so a more shallow IN.
\begin{figure*}[t]
\begin{center}
\includegraphics[width=18cm]{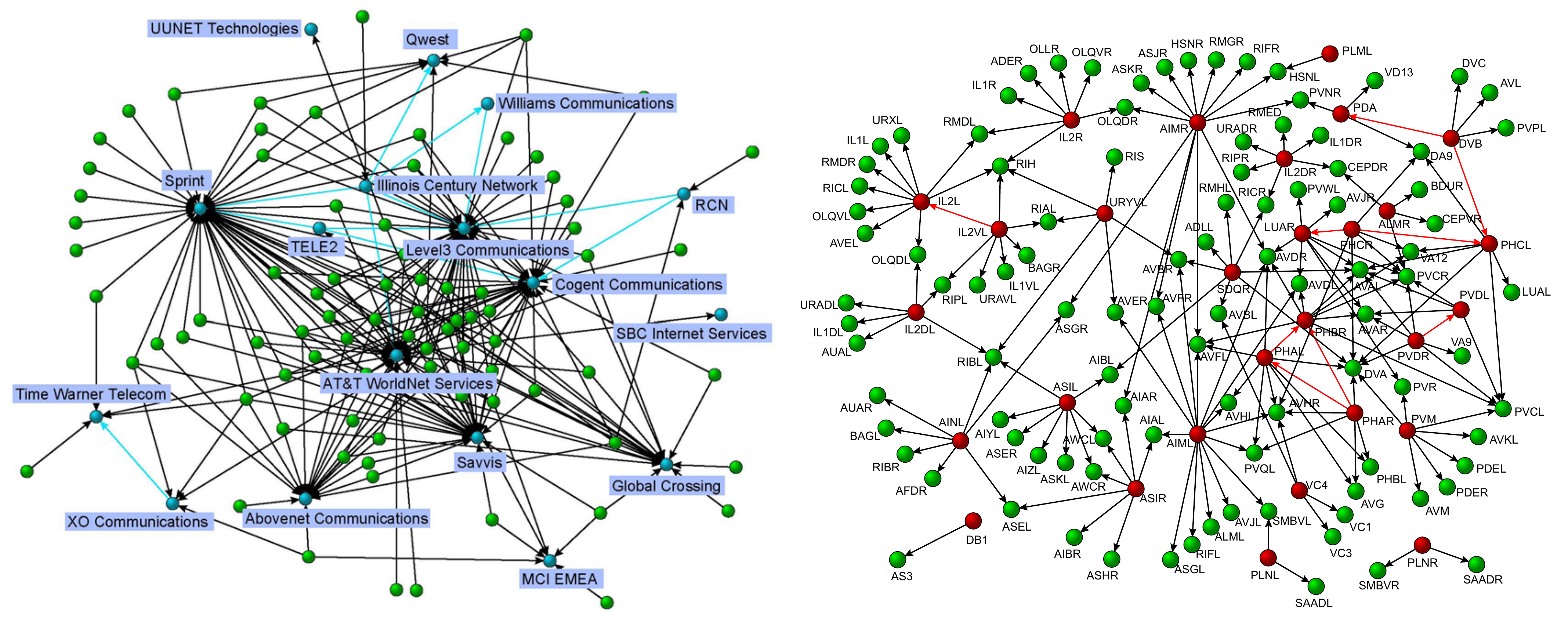}
\vspace{-0.5cm}
\caption{{\bf Left}. Efferent structure in ASR. Edges in blue belong to the OCE and
connect blue nodes within the OUT. {\bf Right}. Afferent structure
in CEN. Edges in red form the ICE and represent connections within red nodes in
the IN. In both representations, nodes in green belong to the core
and edges in black form the corresponding interfaces, OTF and ITF
respectively.} \label{fig:b2}
\end{center}
\end{figure*}

\subsubsection{Synaptic neuronal structure of {\it C.
elegans} (CEN)}
A different family of information transport systems that naturally emerge as
archetypical networks are biological nervous systems. As for most other
complex networks, their structure is intimately related to their
function and the emergent behavior cannot be
understood from the mere summation of the individual neuronal
actions. We focus on the nervous system of
the {\it C. elegans} worm which is practically completely
known~\cite{Chen:2006}.

Network representations of brains display neurons as
vertices and connection between pairs are present whenever a synapse or
gap junction has been observed. We use the updated data
set presented in~\cite{Chen:2006} (http://www.wormatlas.org/).
The pharyngeal
system comprises 20 neurons and is almost totally disconnected from the
rest of the network. It is excluded along unconnected neurons, as
well as connections of the somatic nervous system to non-neural
cells. We further restrict to chemical synapses excluding gap junctions, very different from the
previous in nature and function. For simplicity, polarity or
multiplicity of connections are not taken into account but
directionality is. The synapses are directed in nature but $233$ reciprocal connections has been detected ($12\%$). We
handled this issue by exploiting the imbalance in the number of
observed synapses in each direction, so that we preserve the
directionality of the larger number. In this treatment, just $58$
of them cancel out ($3\%$). The final set contains $279$ nodes
and $1903$ links. As reported previously, it turns out to be a
small-world network~\cite{Watts:1998} with tails of the cumulative
distribution of degrees for both incoming and outgoing neuronal
links that have been reported to be well approximated by exponential
decays~\cite{Amaral:2000}. 

Its percolation layout is surprisingly
close to random organization. In contrast to the Internet, the main
structure consists of a big core with an OUT twice as large as
the IN one (see right graph in Fig.~3 for details about the afferent components), in accordance to the randomized counterpart. The number of edges within the peripheral components is extremely small, so that the C. elegans nervous system seems to rely on clear input and output signals
with direct access to the SCC, the computational processing core, through well populated interfaces.

\section{Interfaces and structural efficiency}
Interfaces play the pivotal role of connecting the IN and OUT components to the network core, the SCC component. Setting aside the discussion of wiring costs~\cite{Achard:2007},
the efficiency of interfaces at fulfilling such task may be loosely defined
using complementary measures able to capture both the amount of load that the
interface edges must bear, the "stress", and the extent of the direct access that peripheral component
nodes have to the SCC through the interface, the "closeness".

\subsection{Stress as random-walk betweenness} As elements transported in the system travel the network, edges
are subject to loads that can be characterized as betweenness,
a topological measure of the number of paths between nodes in
different components that traverse those edges. Betweenness is thus a measure of the
extent to which such edges have control or are under stress because of the flow passing through them. Typically, betweenness is calculated taking into account only
shortest-paths between pairs of nodes~\cite{Freeman:1977}.
Here, we are however interested in more realistic situations and
assume that the topological structure is supporting blind flow
without global knowledge of the system. A more appropriate measure
is therefore the random-walk betweenness~\cite{Newman:2005a}, that
counts all possible routes assuming that information wanders at
random until it finds the target.
Edges with higher random-walk
betweenness are expected to be more important for the spread of
information across the system and, if the load is excessive, bottleneck
effects could even appear.

In order to calculate the random-walk betweenness of the edges at
the interfaces, we slightly modify the original proposal as a
centrality measure for vertices~\cite{Newman:2005a}. The percolation
landscape is explored by means of two symmetric random walks on the
unweighted directed network with homogeneous diffusion probabilities
and absorbing sinks in the nodes of the SCC. Nodes in the IN act as
sources of diffusive particles -either units of energy, packets of
information, economic goods, monetary units...- which spread from
neighbor to neighbor following outgoing links, each chosen with
equal probability among the possibilities. The hopping process is
stopped whenever the diffusive particle arrives to a node in the SCC
following a given link in the ITF, which receives the annotation.
The symmetric process originates particles in the nodes of the OUT,
which travel backwards following incoming links selected with equal
probability among the possibilities, and the diffusion is equally
stopped whenever a node in the SCC is reached through a particular
link in the OTF, which receives the annotation. By repeating the
processes a sufficient number of times for each source node it is
possible to obtain the probability vector that a traveling unit
originated at one of the peripheral components uses the edge $j$ in
the corresponding interface to reach the core. After multiplying by
the size of the source component in number of nodes $N_P$, the
resulting vector $b_{jI}$ informs about the structural load that
each link in the interface supports. Vector $b_{jI}/N_P$ corresponds
to a normalized probability distribution whenever tendrils or tubes
are not considered. The presence of those appendices produce {\it
cul-de-sac} which receive part of the diffusion unloading partially
the interfaces.
\begin{figure}[t]
\begin{center}
\includegraphics[width=8.7cm]{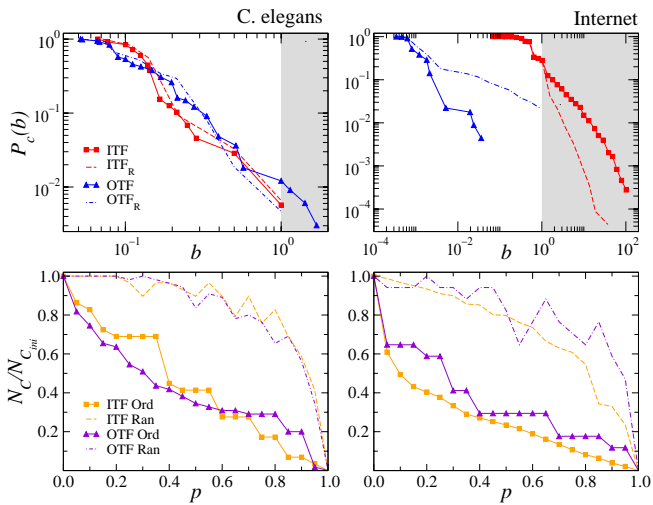}
\vspace{-0.5cm}
\caption{{\bf Upper row panels}. Cumulative random-walk betweenness
distribution for the in- and out-interfaces. The empirical
distributions (symbol lines) are compared against those for the
randomized null models (dashed lines). {\bf Bottom row panels}.
Fraction of nodes remaining in the peripheral components of the real networks after
removing a fraction p of edges of the corresponding interfaces. A
targeted removal by load in decreasing order (symbol lines) is
compared with a random deletion (dashed lines).} \label{fig:b}
\end{center}
\end{figure}

In Fig.~4 (upper panels), we provide the cumulative distribution
$P_c(b)=\sum_{b'>b}P(b')$ of the values $b_{jI}$, the
random-walk betweenness for edge $j$ in interface $I$ (I=ITF or OTF),
which correspond to the loads of the edges
at the interfaces of the ASR and CEN networks. The cumulated probability density function of the loads shows that
they are not uniformly distributed for either network but have heavy tails denoting large fluctuations,
with a few links bearing a much
higher level of structural stress. This heterogeneity is not {\it
per se} indicating that the interface is overstressed. The
random-walk betweenness is moderately highly correlated with
degree~\cite{Newman:2005a} meaning that, in general, vertices with
higher degree tend also to have higher random-walk betweenness, so
that strong disorder in the topology could induce
spurious heterogeneity in the load distribution.

In order to assess whether the structural load could represent a
potential danger of bottleneck formation in traffic related
processes running on the network, one has to define further what is
expected as a low load in the situation of maximal structural
efficiency. We make the
assumption that such efficiency is reached whenever each edge in the interface
carries at most a unitary load. This gives a simple criteria which
makes possible to compare different networks but also different
links of the same interface. At the same time, the results should be
again validated by investigation of the maximally random counterpart. In Fig.~4 (upper panels),
grey areas denote stress regions with loads above $1$. Whereas the CEN network entirely conforms once more to the randomized prediction with the practical totality of loads below the threshold, most edges of the in-interface of the ASR network appear to be overstressed, a clear indicator of the vulnerability of the system. The region of loads much below $1$ usually corresponds to peripheral leaf nodes connected to multiple core nodes.
Apart from a signature of local robustness, this diversification could be interpreted as well as a
quality of being a peripheral spreader or collector of flow.

Finally, the average stress-related structural
efficiency of an interface can be simply approximated as
\begin{equation}
\left< k_B \right>=\frac{E_I}{N_P},
\label{bet}
\end{equation}
that is, the average number of interface edges that mediate between peripheral nodes and the SCC. This average coincides with the inverse of the average betweenness of the edges at the interfaces, $\left< B \right>_{I}=\sum_{j\epsilon F}b_{jI}/E_{I}$. Higher values of $\left<k_B\right>$ are clearly desirable as peripheral nodes would have more routes
to the SCC.

\subsection{Stress and  robustness}
The loads of the edges at the interfaces are
related to their robustness, defined as a measure of the ability of the
interfaces to communicate different components under malfunction or failure.
In the bottom panels of 
Fig.~4, we show the fraction of nodes remaining connected in the peripheral components after the removal of an increasing fraction of edges at the corresponding interface. Two different experiments are performed, the first choosing edges according to load in decreasing order and the second selecting them at random. The results prove that although the interfaces seem to be quite robust against random failures, the failure of high load edges would disconnect a bigger portion of peripheral nodes, thus strongly affecting the behavior of the system. The CEN and AS networks substantially differ in this respect. About 40\% of interface edges must be removed in CEN before 50\% of the peripheral nodes
are disconnected from the SCC in the targeted experiment. By converse, the AS network is more delicate because the same
degree of disconnection is reached by removing just 20\% of the interface edges.
\begin{figure}[t]
\begin{center}
\includegraphics[width=8.7cm]{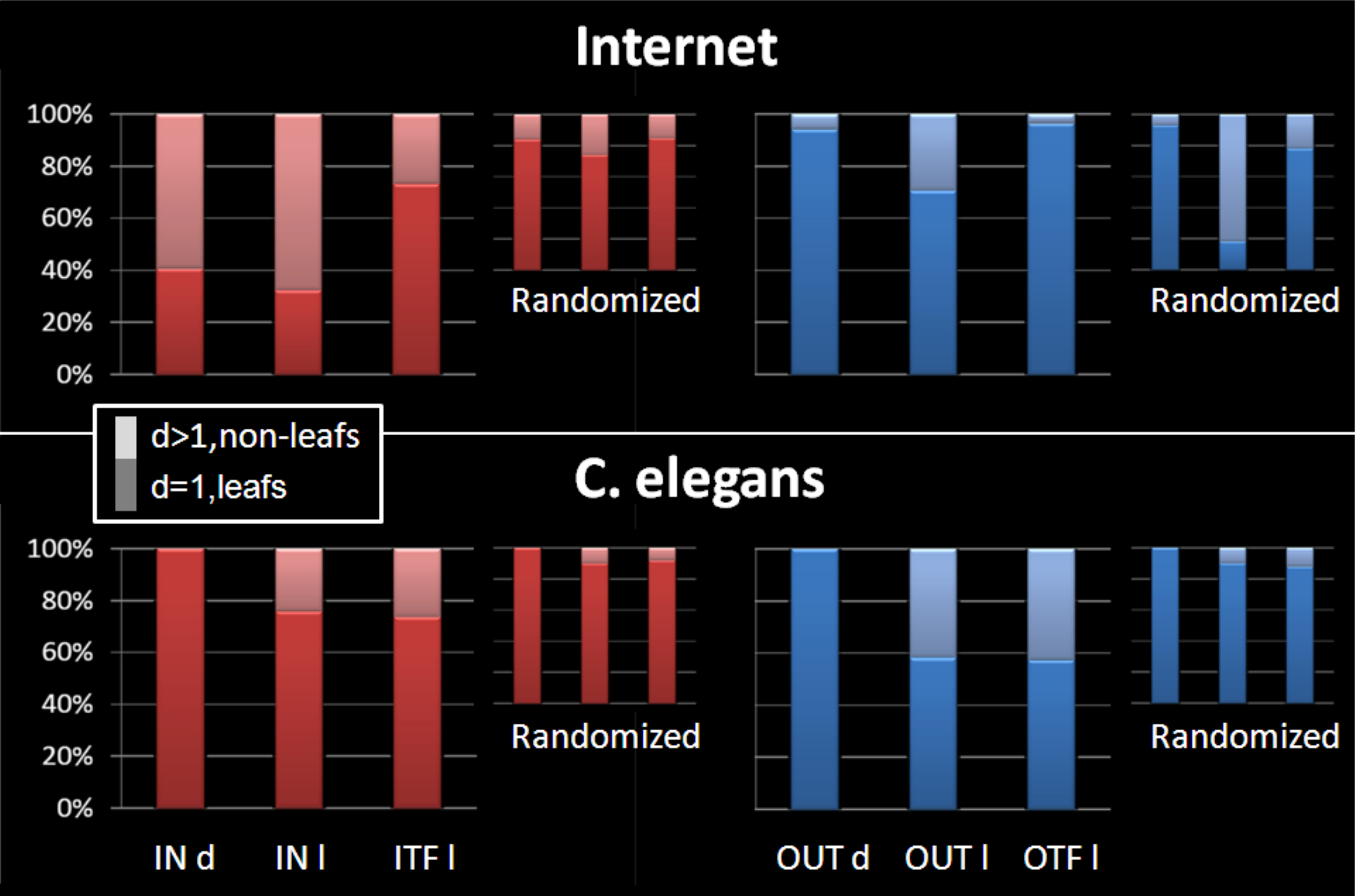}
\vspace{-0.5cm}
\caption{Bar diagrams summarizing
fine details of the peripheral components and the interfaces of ASR
and CEN networks as compared with the randomized null models. Charts in red refer to the afferent structure and those in
blue to the efferent one. The first two bars in each graph refer to the peripheral node components detailing the proportions of nodes at distance $1$ and larger and the proportion of leaf versus non-leaf nodes respectively, and the third bar refers to leaf versus non-leaf edges at the interfaces.} \label{fig:S2}
\end{center}
\end{figure}
\begin{table}[h]
\caption{Interfaces and peripheral components fine details for the ASR and CEN networks and their
randomized counterparts (subscript $R$, average $\pm$ standard deviation rounded off to the first significative figure). Edges at the interfaces and nodes at the components are separated into leafs ($l$) and non-leafs ($nl$), and nodes
directly connected to the core ($d=1$) are distinguished from those at
larger distances ($d>1$).}
\begin{center}
\begin{tabular}{lrr@{$\pm$}lrr@{$\pm$}l}
&Internet &\multicolumn{2}{c}{Internet$_R$}&C. elegans& \multicolumn{2}{c}{C. elegans$_R$}\\
\hline \hline
IN$_{d=1}$& 40.71\%  & 84\% &1\% & 100.00\%   & 100\%&0\%  \\
IN$_{d>1}$& 59.29\%  & 16\%&1\%  & 00.00\%   & 0\% &0\%  \\
IN$_{l}$& 32.34\%  & 74\%&1\%  & 75.86\%   & 90\% &6\% \\
IN$_{nl}$& 67.66\%  & 26\%&1\%  & 24.14\%   & 10\%&6\%  \\
ITF$_{l}$& 73.23\%  & 84.8\%&0.2\%  &  73.71\%& 92\%&5\%\\
ITF$_{nl}$& 26.77\% &  15.2\%&0.2\% & 26.29\% & 8\%&5\%\\
\hline\hline
OUT$_{d=1}$& 94.12\%  & 93\%&4\%  & 100.00\%   & 100\% &0\% \\
OUT$_{d>1}$& 5.88\%  & 7\%&4\%  & 0.00\%   & 0\% &0\% \\
OUT$_{l}$& 70.59\%  &  18\%&3\%  & 58.18\%   & 90\%&6\%  \\
OUT$_{nl}$& 29.41\%  &  80\%&3\%  & 41.82\%   & 10\%&6\%  \\
OTF$_{l}$ & 96.46\%  & 78\%&9\%  &  57.27\%& 88\%&7\%\\
OTF$_{nl}$& 3.54\% &  22\%&9\% &  42.73\% & 12\%&7\%\\
\hline\hline
\end{tabular}
\end{center}
\label{table_3}
\end{table}

\subsection{Closeness}
The random walk methodology presented above cannot discriminate
between peripheral conformations with different access to the SCC if equal loads are associated
to interface edges (as a simple example, see tree-like groups $A$ and $B$ in
Fig.1). The concept of closeness allows us to shed light on the different efficiencies that characterize these dissimilar architectures.

By convention, leaf vertices are those with in-degree $0$ or out-degree $0$, so that they are restricted to belong to a peripheral component.
In-leaf edges (out-leaf edges) are considered as directed links leaving
from (pointing to) a leaf vertex\footnote{Strictly speaking, vertices with in-degree $0$ are usually referred as root vertices. We refer to them as leaf vertices for economy of language. Note also that according to the definitions of in-leaf (out-leaf) edges, the in (out) interface cannot contain out-leaf (in-leaf) edges and that from the perspective of vertices the definitions would be reversed.}.
Non-leaf edges in the interfaces are
the ones that ensure the communication from/to nodes not directly
connected to the core. These non-leaf edges are the potentially
responsible for  bottleneck effects, since they service from more than a single IN or OUT node.
A first estimate of how this topological considerations affects efficiency at the structural level is given by the closeness average degree,
\begin{equation}
\left< k_{C} \right>=\frac{E_{I,nl}}{N_{P,d>1}},  \label{ke}
\end{equation}
which is the number of interface non-leaf edges available for each peripheral node which is not
directly connected to the SCC (thus, with a distance $d$ from the SCC greater than $1$).
\begin{table}[t]
\caption{Structural efficiency average degrees for ASR and CEN and their
randomized counterparts (subscript $R$, values are average $\pm$ standard deviation rounded off to the first significative figure). Infinite closeness averages come from the fact of all peripheral nodes being directly connected to the core.}
\begin{center}
\begin{tabular}{lrr@{$\pm$}lrr@{$\pm$}l}
&ASR &\multicolumn{2}{c}{ASR$_R$}&CEN&\multicolumn{2}{c}{ CEN$_R$}\\
\hline \hline
$\left< k_B\right>_{IN}$& 0.54 &    1.15&0.04     & 6.03    &  6.3&0.3 \\
$\left< k_{C}\right>_{IN}$& 0.24 &   1.1&0.1     & $\infty$     &  $\infty$ &0 \\
\hline\hline
$\left< k_B\right>_{OUT}$& 13.29 &    8&1     & 6.00    &  6.5&0.2 \\
$\left< k_{C}\right>_{OUT}$& 8.00 &   30&20     & $\infty$    &  $\infty$&0 \\
\hline\hline
\end{tabular}
\end{center}
\label{table_2}
\end{table}

Values for the decomposition of the ASR and CEN interfaces and the peripheral
components into leafs and non-leaf units along with average degree
efficiency measures as defined in
Eq.~(\ref{bet}) and Eq.~(\ref{ke}) are shown in
Table~2, Table~3, and Fig.~5. In general terms, the higher the averages the more
structurally efficient the system is expected to be. An important imbalance is observed between the in and out values for ASR. According to the average values, the in-interface presents a certain level of inefficiency, with low average degrees combined with a low number of leafs, much below random expectations. In this situation, potential bottleneck effects are more likely. In contrast, the out component shows high levels of structural efficiency, with the practical totality of nodes being root nodes directly connected to the core. On the other hand, all peripheral CEN nodes have direct access to the core, a signature of high efficiency.

\subsection{Maximum structural efficiency and the ``hairy'' ball}
Under the requirements of low stress and high closeness, and in the approximation of inexpensive
edges, maximum efficiency would be
realized by a percolation landscape structured
as a perfect ``hairy ball'', with all the nodes in the peripheral components
directly attached to the core through leaf edges, each carrying at most a unitary load
\footnote{The absence of nodes at
distances larger than $1$ could involve a marginal deviation
from the ``hairy ball'' conformation with a few loads slightly
greater than $1$ due to inner connections in the peripheral
component.}, thus without endangering bottleneck effects.
Moreover, the interfaces would be robust because the
failure or malfunctioning of any of its edges would affect a minimum
number of nodes in the peripheral components. Finally, all peripheral nodes
would have direct access to the core. Any departure from the ``hairy
ball'' paradigm would lead to situations in which at least one of
the two or both requests for structural efficiency, low stress and
high closeness, are violated to some degree.

\section{Conclusions}
Our thorough analysis of percolation landscapes shows that the conformation of interfaces plays a central role in the performance of complex flow networks as global transport systems,
governing their efficiency against bottlenecks and their robustness against failures.
We highlight that, from the purely structural efficiency perspective, a "hairy ball" design would be optimal.
Appealingly, such behavior may be even displayed by a very close to random architecture as seen for the synaptic neuronal network of C. elegans. Of the two real systems analyzed in this work, this is the one much closer to such optimality whereas the Internet network presents inefficiencies. These findings point to two, not mutually exclusive, interpretations. On the one
hand, different adaptation
dynamics are surely at work: whereas the present structure of the C. elegans nervous system
tries to optimize its {\it collective} performance without inter-neuron competition, the Internet network
emerges, due to its customer-provider relations, as a competitive network where it is not
the global optimization which is sought for but rather the individual Internet service provider gain. In this respect,
global efficiency is important only in relation to its marketable value. Interestingly, the Internet customer-provider network outperforms its randomized version in the OTF and OUT components, which describe the
ultimate cash flow, and underperforms it in the afferent components, where end-users are. On the other hand, evolution of the worm nervous system might have allowed better architectures to emerge, due to its evolutionary time-scale (hundred of millions of years) running much longer than the time-span of existence of the commercial Internet network (slightly more than ten years).

Clearly, these results only shed light on the basic structural ingredients for efficiency and robustness.
Indeed, several other constraints ({\it e.g.} costs of edge deployment and maintenance or capacity), are at play
which should be taken into account for more precise and system specific analysis. Yet,
percolation landscapes represent a first general framework to highlight potential problems
in a network structure, possibly suggesting specific actions to reinforce stressed elements or the redistribution of loads so to reduce the risk of bottlenecks and the impacts of failures.

\begin{acknowledgments}
The authors thank Mari{\'a}n Bogu{\~n}{\'a} for useful comments and
discussions. This work has been financially supported by DELIS under
contract FET Open 001907 and the SER-Bern under contract 02.0234.
\end{acknowledgments}


\begin{thebibliography}{26}
\expandafter\ifx\csname natexlab\endcsname\relax\def\natexlab#1{#1}\fi
\expandafter\ifx\csname bibnamefont\endcsname\relax
  \def\bibnamefont#1{#1}\fi
\expandafter\ifx\csname bibfnamefont\endcsname\relax
  \def\bibfnamefont#1{#1}\fi
\expandafter\ifx\csname citenamefont\endcsname\relax
  \def\citenamefont#1{#1}\fi
\expandafter\ifx\csname url\endcsname\relax
  \def\url#1{\texttt{#1}}\fi
\expandafter\ifx\csname urlprefix\endcsname\relax\def\urlprefix{URL }\fi
\providecommand{\bibinfo}[2]{#2}
\providecommand{\eprint}[2][]{\url{#2}}

\bibitem[{\citenamefont{Albert and Barab{\'a}si}(2002)}]{Albert:2002}
\bibinfo{author}{\bibfnamefont{R.}~\bibnamefont{Albert}} \bibnamefont{and}
  \bibinfo{author}{\bibfnamefont{A.-L.} \bibnamefont{Barab{\'a}si}},
  \bibinfo{journal}{Rev. Mod. Phys.} \textbf{\bibinfo{volume}{74}},
  \bibinfo{pages}{47} (\bibinfo{year}{2002}).

\bibitem[{\citenamefont{Dorogovtsev and Mendes}(2003)}]{Dorogovtsev:2003}
\bibinfo{author}{\bibfnamefont{S.~N.} \bibnamefont{Dorogovtsev}}
  \bibnamefont{and} \bibinfo{author}{\bibfnamefont{J.~F.~F.}
  \bibnamefont{Mendes}}, \emph{\bibinfo{title}{Evolution of networks: From
  biological nets to the Internet and WWW}} (\bibinfo{publisher}{Oxford
  University Press}, \bibinfo{address}{Oxford}, \bibinfo{year}{2003}).

\bibitem[{\citenamefont{Newman}(2003)}]{Newman:2003}
\bibinfo{author}{\bibfnamefont{M.~E.~J.} \bibnamefont{Newman}},
  \bibinfo{journal}{SIAM Review} \textbf{\bibinfo{volume}{45}},
  \bibinfo{pages}{167} (\bibinfo{year}{2003}).

\bibitem[{\citenamefont{Broder et~al.}(2000)\citenamefont{Broder, Kumar,
  Maghoul, Raghavan, Rajagopalan, Stata, Tomkins, and Wiener}}]{Broder:2000}
\bibinfo{author}{\bibfnamefont{A.}~\bibnamefont{Broder}},
  \bibinfo{author}{\bibfnamefont{R.}~\bibnamefont{Kumar}},
  \bibinfo{author}{\bibfnamefont{F.}~\bibnamefont{Maghoul}},
  \bibinfo{author}{\bibfnamefont{P.}~\bibnamefont{Raghavan}},
  \bibinfo{author}{\bibfnamefont{S.}~\bibnamefont{Rajagopalan}},
  \bibinfo{author}{\bibfnamefont{S.}~\bibnamefont{Stata}},
  \bibinfo{author}{\bibfnamefont{A.}~\bibnamefont{Tomkins}}, \bibnamefont{and}
  \bibinfo{author}{\bibfnamefont{J.}~\bibnamefont{Wiener}},
  \bibinfo{journal}{Computer Networks} \textbf{\bibinfo{volume}{33}},
  \bibinfo{pages}{309} (\bibinfo{year}{2000}).

\bibitem[{\citenamefont{Guimer\`{a} et~al.}(2007)\citenamefont{Guimer\`{a},
  Sales-Pardo, and Amaral}}]{Guimera:2007}
\bibinfo{author}{\bibfnamefont{R.}~\bibnamefont{Guimer\`{a}}},
  \bibinfo{author}{\bibfnamefont{M.}~\bibnamefont{Sales-Pardo}},
  \bibnamefont{and} \bibinfo{author}{\bibfnamefont{L.~A.~N.}
  \bibnamefont{Amaral}}, \bibinfo{journal}{Nature Physics}
  \textbf{\bibinfo{volume}{3}}, \bibinfo{pages}{63} (\bibinfo{year}{2007}).

\bibitem[{\citenamefont{Latora and Marchiori}(2001)}]{Latora:2001}
\bibinfo{author}{\bibfnamefont{V.}~\bibnamefont{Latora}} \bibnamefont{and}
  \bibinfo{author}{\bibfnamefont{M.}~\bibnamefont{Marchiori}},
  \bibinfo{journal}{Phys. Rev. Lett.} \textbf{\bibinfo{volume}{87}},
  \bibinfo{pages}{198701} (\bibinfo{year}{2001}).

\bibitem[{\citenamefont{S.~Sreenivasan and Stanley}(2007)}]{Sreenivasan:2007}
\bibinfo{author}{\bibfnamefont{E.~L. Z.~T.} \bibnamefont{S.~Sreenivasan},
  \bibfnamefont{R.~Cohen}} \bibnamefont{and}
  \bibinfo{author}{\bibfnamefont{H.~E.} \bibnamefont{Stanley}},
  \bibinfo{journal}{Physical Review E} \textbf{\bibinfo{volume}{75}},
  \bibinfo{pages}{036105} (\bibinfo{year}{2007}).

\bibitem[{\citenamefont{Gallos et~al.}(2007)\citenamefont{Gallos, Song, Havlin,
  and Makse}}]{Gallos:2007}
\bibinfo{author}{\bibfnamefont{L.~K.} \bibnamefont{Gallos}},
  \bibinfo{author}{\bibfnamefont{C.}~\bibnamefont{Song}},
  \bibinfo{author}{\bibfnamefont{S.}~\bibnamefont{Havlin}}, \bibnamefont{and}
  \bibinfo{author}{\bibfnamefont{H.~A.} \bibnamefont{Makse}},
  \bibinfo{journal}{Proc. Natl. Acad. Sci. USA} \textbf{\bibinfo{volume}{104}},
  \bibinfo{pages}{7746–7751} (\bibinfo{year}{2007}).

\bibitem[{\citenamefont{Fischer and Sauer}(2005)}]{Fisher:2005}
\bibinfo{author}{\bibfnamefont{E.}~\bibnamefont{Fischer}} \bibnamefont{and}
  \bibinfo{author}{\bibfnamefont{U.}~\bibnamefont{Sauer}},
  \bibinfo{journal}{Nature Genetics} \textbf{\bibinfo{volume}{37}},
  \bibinfo{pages}{636} (\bibinfo{year}{2005}).

\bibitem[{\citenamefont{Csete and Doyle}(2004)}]{Csete:2004}
\bibinfo{author}{\bibfnamefont{M.}~\bibnamefont{Csete}} \bibnamefont{and}
  \bibinfo{author}{\bibfnamefont{J.}~\bibnamefont{Doyle}},
  \bibinfo{journal}{TRENDS in Biotechnology} \textbf{\bibinfo{volume}{22}},
  \bibinfo{pages}{446} (\bibinfo{year}{2004}).

\bibitem[{\citenamefont{Segr\`{e} et~al.}(2002)\citenamefont{Segr\`{e}, Vitkup,
  and Church}}]{Segre:2002}
\bibinfo{author}{\bibfnamefont{D.}~\bibnamefont{Segr\`{e}}},
  \bibinfo{author}{\bibfnamefont{D.}~\bibnamefont{Vitkup}}, \bibnamefont{and}
  \bibinfo{author}{\bibfnamefont{G.~M.} \bibnamefont{Church}},
  \bibinfo{journal}{Proc. Natl. Acad. Sci. USA} \textbf{\bibinfo{volume}{99}},
  \bibinfo{pages}{15112} (\bibinfo{year}{2002}).

\bibitem[{\citenamefont{Newman et~al.}(2001)\citenamefont{Newman, Strogatz, and
  Watts}}]{Newman:2001b}
\bibinfo{author}{\bibfnamefont{M.~E.~J.} \bibnamefont{Newman}},
  \bibinfo{author}{\bibfnamefont{S.~H.} \bibnamefont{Strogatz}},
  \bibnamefont{and} \bibinfo{author}{\bibfnamefont{D.~J.} \bibnamefont{Watts}},
  \bibinfo{journal}{Phys. Rev. E} \textbf{\bibinfo{volume}{64}},
  \bibinfo{pages}{026118} (\bibinfo{year}{2001}).

\bibitem[{\citenamefont{Serrano and Rios}(2007)}]{Serrano:2007c}
\bibinfo{author}{\bibfnamefont{M.~A.} \bibnamefont{Serrano}} \bibnamefont{and}
  \bibinfo{author}{\bibfnamefont{P.~D.~L.} \bibnamefont{Rios}},
  \bibinfo{journal}{arXiv:0706.3156v1 [cond-mat.dis-nn]}
  (\bibinfo{year}{2007}).

\bibitem[{\citenamefont{Bogu{\~n}\'{a} and Serrano}(2005)}]{Boguna:2005}
\bibinfo{author}{\bibfnamefont{M.}~\bibnamefont{Bogu{\~n}\'{a}}}
  \bibnamefont{and} \bibinfo{author}{\bibfnamefont{M.~A.}
  \bibnamefont{Serrano}}, \bibinfo{journal}{Phys. Rev. E}
  \textbf{\bibinfo{volume}{72}}, \bibinfo{pages}{016106}
  (\bibinfo{year}{2005}).

\bibitem[{\citenamefont{Dorogovtsev et~al.}(2001)\citenamefont{Dorogovtsev,
  Mendes, and Samukhin}}]{Dorogovtsev:2001}
\bibinfo{author}{\bibfnamefont{S.~N.} \bibnamefont{Dorogovtsev}},
  \bibinfo{author}{\bibfnamefont{J.~F.~F.} \bibnamefont{Mendes}},
  \bibnamefont{and} \bibinfo{author}{\bibfnamefont{A.~N.}
  \bibnamefont{Samukhin}}, \bibinfo{journal}{Phys. Rev. E}
  \textbf{\bibinfo{volume}{64}}, \bibinfo{pages}{066110}
  (\bibinfo{year}{2001}).

\bibitem[{\citenamefont{Newman}(2002)}]{Newman:2002a}
\bibinfo{author}{\bibfnamefont{M.~E.~J.} \bibnamefont{Newman}},
  \bibinfo{journal}{Phys. Rev. Lett.} \textbf{\bibinfo{volume}{89}},
  \bibinfo{pages}{208701} (\bibinfo{year}{2002}).

\bibitem[{\citenamefont{Pastor-Satorras and
  Vespignani}(2004)}]{RomusVespasbook}
\bibinfo{author}{\bibfnamefont{R.}~\bibnamefont{Pastor-Satorras}}
  \bibnamefont{and}
  \bibinfo{author}{\bibfnamefont{A.}~\bibnamefont{Vespignani}},
  \emph{\bibinfo{title}{Evolution and Structure of the Internet. A Statistical
  Physics Approach}} (\bibinfo{publisher}{Cambridge University Press},
  \bibinfo{address}{Cambridge}, \bibinfo{year}{2004}).

\bibitem[{\citenamefont{Dimitropoulos et~al.}(2007)\citenamefont{Dimitropoulos,
  Krioukov, Fomenkov, Huffaker, Hyun, kc~claffy, and
  Riley}}]{Dimitropoulos:2007}
\bibinfo{author}{\bibfnamefont{X.}~\bibnamefont{Dimitropoulos}},
  \bibinfo{author}{\bibfnamefont{D.}~\bibnamefont{Krioukov}},
  \bibinfo{author}{\bibfnamefont{M.}~\bibnamefont{Fomenkov}},
  \bibinfo{author}{\bibfnamefont{B.}~\bibnamefont{Huffaker}},
  \bibinfo{author}{\bibfnamefont{Y.}~\bibnamefont{Hyun}},
  \bibinfo{author}{\bibnamefont{kc~claffy}}, \bibnamefont{and}
  \bibinfo{author}{\bibfnamefont{G.}~\bibnamefont{Riley}},
  \bibinfo{journal}{ACM SIGCOMM Computer Communication Review}
  \textbf{\bibinfo{volume}{37}}, \bibinfo{pages}{29} (\bibinfo{year}{2007}).

\bibitem[{\citenamefont{Chen et~al.}(2006)\citenamefont{Chen, Hall, and
  Chklovskii}}]{Chen:2006}
\bibinfo{author}{\bibfnamefont{B.~L.} \bibnamefont{Chen}},
  \bibinfo{author}{\bibfnamefont{D.~H.} \bibnamefont{Hall}}, \bibnamefont{and}
  \bibinfo{author}{\bibfnamefont{D.~B.} \bibnamefont{Chklovskii}},
  \bibinfo{journal}{Proc. Natl. Acad. Sci. USA} \textbf{\bibinfo{volume}{103}},
  \bibinfo{pages}{4723} (\bibinfo{year}{2006}).

\bibitem[{\citenamefont{Achard and Bullmore}(2007)}]{Achard:2007}
\bibinfo{author}{\bibfnamefont{S.}~\bibnamefont{Achard}} \bibnamefont{and}
  \bibinfo{author}{\bibfnamefont{E.}~\bibnamefont{Bullmore}},
  \bibinfo{journal}{PLOS Computational Biology} \textbf{\bibinfo{volume}{3}},
  \bibinfo{pages}{174} (\bibinfo{year}{2007}).

\bibitem[{\citenamefont{Freeman}(1977)}]{Freeman:1977}
\bibinfo{author}{\bibfnamefont{L.}~\bibnamefont{Freeman}},
  \bibinfo{journal}{Sociometry} \textbf{\bibinfo{volume}{40}},
  \bibinfo{pages}{35–41} (\bibinfo{year}{1977}).

\bibitem[{\citenamefont{Newman}(2005)}]{Newman:2005a}
\bibinfo{author}{\bibfnamefont{M.~E.~J.} \bibnamefont{Newman}},
  \bibinfo{journal}{Social Networks} \textbf{\bibinfo{volume}{27}},
  \bibinfo{pages}{39} (\bibinfo{year}{2005}).

\bibitem[{\citenamefont{Watts and Strogatz}(1998)}]{Watts:1998}
\bibinfo{author}{\bibfnamefont{D.~J.} \bibnamefont{Watts}} \bibnamefont{and}
  \bibinfo{author}{\bibfnamefont{S.~H.} \bibnamefont{Strogatz}},
  \bibinfo{journal}{Nature} \textbf{\bibinfo{volume}{393}},
  \bibinfo{pages}{440} (\bibinfo{year}{1998}).

\bibitem[{\citenamefont{Amaral et~al.}(2000)\citenamefont{Amaral, Scala,
  Barth\'{e}lemy, and Stanley}}]{Amaral:2000}
\bibinfo{author}{\bibfnamefont{L.~A.~N.} \bibnamefont{Amaral}},
  \bibinfo{author}{\bibfnamefont{A.}~\bibnamefont{Scala}},
  \bibinfo{author}{\bibfnamefont{M.}~\bibnamefont{Barth\'{e}lemy}},
  \bibnamefont{and} \bibinfo{author}{\bibfnamefont{H.~E.}
  \bibnamefont{Stanley}}, \bibinfo{journal}{Proc. Natl. Acad. Sci. USA}
  \textbf{\bibinfo{volume}{97}}, \bibinfo{pages}{11149} (\bibinfo{year}{2000}).

\bibitem[{\citenamefont{Maslov and Sneppen}(2002)}]{Maslov:2002}
\bibinfo{author}{\bibfnamefont{S.}~\bibnamefont{Maslov}} \bibnamefont{and}
  \bibinfo{author}{\bibfnamefont{K.}~\bibnamefont{Sneppen}},
  \bibinfo{journal}{Science} \textbf{\bibinfo{volume}{296}},
  \bibinfo{pages}{910–913} (\bibinfo{year}{2002}).

\bibitem[{\citenamefont{Bogu{\~n}\'{a}
  et~al.}(2004)\citenamefont{Bogu{\~n}\'{a}, Pastor-Satorras, and
  Vespignani}}]{Boguna:2004}
\bibinfo{author}{\bibfnamefont{M.}~\bibnamefont{Bogu{\~n}\'{a}}},
  \bibinfo{author}{\bibfnamefont{R.}~\bibnamefont{Pastor-Satorras}},
  \bibnamefont{and}
  \bibinfo{author}{\bibfnamefont{A.}~\bibnamefont{Vespignani}},
  \bibinfo{journal}{European Physical Journal B} \textbf{\bibinfo{volume}{38}},
  \bibinfo{pages}{205} (\bibinfo{year}{2004}).

\end{thebibliography}
\end{document}